\documentclass[10pt,a4paper]{article}
\usepackage[utf8]{inputenc}
\usepackage{amsmath}
\usepackage{amsfonts}
\usepackage{amssymb}
\usepackage{graphicx}
\usepackage{authblk}
\usepackage{tcolorbox}
\usepackage[title]{appendix}
\usepackage[normalem]{ulem}
\title{Unveiling the relation between herding and liquidity with trader lead-lag networks}
\author[1]{Carlo Campajola}
\affil[1]{Scuola Normale Superiore di Pisa}
\author[2]{Fabrizio Lillo}
\affil[2]{University of Bologna - Department of Mathematics}
\author[2]{Daniele Tantari}
\begin{document}
\maketitle

\begin{abstract}
    We propose a method to infer lead-lag networks of traders from the observation of their trade record as well as to reconstruct their state of supply and demand when they do not trade. The method relies on the Kinetic Ising model to describe how information propagates among traders, assigning a positive or negative ``opinion" to all agents about whether the traded asset price will go up or down. This opinion is reflected by their trading behavior, but whenever the trader is not active in a given time window, a missing value will arise. Using a recently developed inference algorithm, we are able to reconstruct a lead-lag network and to estimate the unobserved opinions, giving a clearer picture about the state of supply and demand in the market at all times.
    We apply our method to a dataset of clients of a major dealer in the Foreign Exchange market at the 5 minutes time scale.  We identify leading players in the market and define a herding measure based on the observed and inferred opinions. We show the causal link between herding and liquidity in the inter-dealer market used by dealers to rebalance their inventories.
\end{abstract}
\section{Introduction}
A significant part of risk management for financial intermediaries is related to the mitigation of the adverse selection risk \cite{Kyle,GM}, namely the risk of trading with a counterpart that has access to better information on the traded asset. This risk is exacerbated in contexts where multiple counterparties might - consciously or not - coordinate their trades, introducing not only an adverse selection risk against a specific counterpart but against a group of traders, a situation typically referred to as inventory risk \cite{ho1980dealer}. Understanding how information propagates in the market is crucial to identify key players that can forerun the order flow, a knowledge that an intermediary can exploit to better hedge against inventory risk.

Methods to detect lead-lag relationships between financial variables have been extensively studied in the literature, starting with correlations between financial assets \cite{jegadeesh1995overreaction} and evolving towards more complex measurement methods such as cascade models \cite{lux2001turbulence} or Vector AutoRegressive models \cite{barigozzi2019nets}. In recent years there has been a rising interest in methods to cluster together traders based on their strategic and behavioral features as well as studying how they influence each other. A prominent example is the Statistically Validated Networks (SVN) methodology, first described in Tumminello et al. \cite{tumminello2011statistically} and then applied to financial data \cite{tumminello2012identification, curme2015emergence, musciotto2018long}, which has then been extended to the Statistically Validated Lead-Lag Networks methodology \cite{challet2018statistically, cordi2019market} to analyse how investors can be classified based on their strategic behaviour and which clusters correlate at different time-scales. Challet et al. \cite{challet2016trader} proposed a Machine Learning method to construct lead-lag networks between clusters of investors and predict the order flow, while Guti{\'e}rrez-Roig et al. \cite{Gutierrez-Roig2019} rely on information-based methods to achieve similar results.

We propose our approach as an alternative to the aforementioned methods, introducing the Kinetic Ising Model as an opinion spreading mechanism whose parameters can be inferred from the data, as shown in a recent methodological article we published \cite{campajola2019inference}. 

In the original formulation of the Ising model \cite{ising1925beitrag}, a set of binary variables called \textit{spins} represent magnetic dipoles of atoms in a magnetic material, which can be oriented along an axis taking positive or negative value. Their electromagnetic interaction is modelled by a \textit{coupling} factor, which, in the simplest version, is assumed to be equal for all spins and in more complicated versions can be a matrix of pair-specific factors. The Kinetic Ising Model is the out-of-equilibrium implementation of such a model, defined by approximating in discrete time the appropriate set of stochastic differential equations that describe the dynamics of the Ising model in continuous time. The resulting model is a simple yet effective description of the time evolution of a set of binary random variables where, like in Vector AutoRegressive (VAR) models, there is a lagged interaction, meaning that their probability distribution function at time $t$ depends on their and others values at time $t-1$ through a logistic link function.

The goal is to find significant lead-lag relationships between single market participants on the intra-day time-scale, as well as exploiting these lead-lag relations to estimate the current implicit state of supply and demand. There are two main innovations in our approach with respect to the above mentioned ones: on the one hand, treating the data as a whole in a multivariate model, instead of running multiple pairwise tests - as previously cited methods do - allows us to correctly identify correlations and causalities, whereas a pairwise approach is potentially prone to cases where spurious effects appear; on the other hand, we also have the ability to handle missing observations, which in the case of financial markets is an effect of the intrinsic asynchronicity of trade records \cite{ait2010high, Corsi2012}.

The main purpose of financial markets is to aggregate the public opinion about a particular asset, determining the correct price as the optimal match between supply and demand. The opinion of a particular trader about the asset price is thus expressed when they perform a transaction: when they buy an asset at price $p$, they believe the correct price (the ``value" of the asset) is $p' > p$, and vice-versa. Due to transaction costs, limited liquidity and other frictional effects, the traders incur in a cost whenever they want to express their opinion, inducing them to trade less than they would in an ideal situation. As a result, when looking at trade records on the intra-day time-scale, it is very hard to aggregate time at a level such that every participant trades in every time slice. However it is reasonable to assume that, even if a trader has not traded in the last time interval, they still hold an opinion about the asset, which could be reflected in other trades they perform on other markets or could influence other traders in their future actions.

We choose to model this system through the Kinetic Ising Model, assuming traders' opinions can either be positive (belief that $p' > p$) or negative ($p' < p$) and thus be represented by binary spins that evolve in discrete time. Their coupling factors will then carry the information about lead-lag relationships in the spreading of opinions at the considered time-scale. As mentioned, the only observations available about such opinions are the trades that investors make, meaning that the data will likely present a significant amount of missing values if one takes a reasonably short time step. A good reason to choose the Kinetic Ising Model then is the possibility to infer the model parameters efficiently even from incomplete data, thanks to the Expectation-Maximization-like algorithm that we recently developed \cite{campajola2019inference}, while also getting a Maximum Likelihood estimate of the unobserved opinions. The intuition behind the algorithm is that, given the model parameters, one can analytically take expectations on the missing values, and then use such expectations to improve the inference of the model parameters themselves. Thus, by recurring this two-step procedure until a maximum of the log-likelihood is reached, one jointly estimates the model parameters \textit{and} the expected values of unknown opinions. Such expectations can then be used to make an informed guess about the hidden opinion, by simply taking their sign.

The case is particularly relevant for the foreign exchange (FX) market. The market has a multi-dealer organization, where a centralized double-auction exchange is accessible to few market members (the dealers) which in turn offer, through their proprietary platform, a trading service to their clients. The dealer then acts as a liquidity provider, while also absorbing temporary shocks in supply and demand through its inventory which is then rebalanced by trading with other dealers on the centralized platform. Optimal dealership (mostly known as optimal market making) is a vastly studied problem in finance (see Gu{\'e}ant \cite{gueant2016financial} for a comprehensive review), trying to devise how to optimally rebalance the inventory one accumulates when satisfying clients' requests and what is the fee the dealer has to charge clients in exchange for the immediacy of their transaction. One of the costs faced by dealers is the cost of liquidity on the inter-dealer market, which can be particularly high when all market participants experience the same kind of pressure from their clients. To predict what this cost will be it can be useful to understand what the aggregate opinion of traders is, even the ones the dealer doesn't observe due to lack of trading activity, either because they might influence other clients actions or because they are active with other competing dealers and will eventually impact the cost of liquidity shortly afterwards.

Our modelling approach also allows to analyse the inferred lead-lag networks to identify key nodes in the opinion spreading process, whether the network changes over time as traders enter and exit the market, and to study how influential nodes are relevant for the prediction of the order flow and future liquidity.

The paper is organized as follows: in Section \ref{sec:method} we briefly describe the Kinetic Ising Model and the inference algorithm, in Section \ref{sec:data} we describe the dataset we use, in Section \ref{sec:results} we show the results coming from multiple network analysis metrics (Sections \ref{sec:influencers} and \ref{sec:neighb}), we analyze the performance of the model when trying to forecast the order flow (Section \ref{sec:OOS}) and we define a herding measure from the inferred opinions, for which we test Granger Causality \cite{granger1969investigating} effects with several liquidity imbalance measures (Section \ref{sec:GC}). Section \ref{sec:concl} concludes the paper.

\section{Method}\label{sec:method}

The method we adopt relies on the Kinetic Ising Model (KIM) \cite{derrida1987exactly,crisanti1988dynamics}, a Markovian model describing the dynamics of interacting binary random variables through the Boltzmann distribution, which belongs to the exponential family. The model has been developed in the physics literature as the out-of-equilibrium version of the popular Ising Model \cite{ising1925beitrag,brush1967history}, originally intended to describe the physics of ferromagnets. However, thanks to its simplicity and rich behaviour, it lends itself to be used in other contexts such as neuroscience \cite{cocco2017functional}, computational biology \cite{tanaka1977model,imparato2007ising,agliari2011thermodynamic}, economics and finance \cite{Bouchaud2013, SornetteReview} and even machine learning with neural networks \cite{lecun2015deep,hornik1989multilayer}.

The model itself belongs to the family of logistic regression models, in this particular case in a multivariate and autoregressive form, which we briefly describe in the following and refer to the literature for more accurate treatments.

Consider a $N$-dimensional vector of binary random variables $y \in \lbrace -1, +1, \rbrace^N$ evolving in time $t = 1,\dots,T$, representing, in our case, the (positive or negative) opinion traders hold about the traded asset at each time step. We want to model the spreading of these opinions, and we thus define the transition probability
\begin{equation}
    p\left[y(t+1) \vert y(t) \right] = Z^{-1} (t) \exp \left[ \sum_{\langle i,j \rangle} y_i(t+1) J_{ij} y_j(t) + \sum_i y_i(t+1)  \left( h_i + b_i r(t) \right) \right]
    \label{eq:KIM}
\end{equation}
where $\langle i,j \rangle$ is the count of neighbouring pairs of traders on an underlying network, $J_{ij}$ are the coupling parameters between trader $i$'s opinion at time $t+1$ and trader $j$'s at time $t$, $h$ is the vector of agent-specific bias parameters, $b$ is the vector coupling agent opinions to the log-returns on the asset price $r(t)$\footnote{Or any other external regressor} and $Z(t)$ is a normalizing constant, also known as the partition function.

In the specific setting we consider here, we treat observed trade signs $s_i(t)$ as observations of trader $i$'s opinion $y_i(t)$, implying that traders not active at time $t$, say agent $a$, still hold an opinion, which is hidden to the market and is thus handled as a missing observation we call $\sigma_a(t)$. The implicit mapping between the opinion dynamics and the observations is given by a set of right-invertible matrices $G(t) \in \lbrace 0,1 \rbrace^{M(t) \times N}$, where $M(t)$ is the number of observed trades at time $t$, such that $s(t) = G(t) y(t)$. The specific choice of $G(t)$ in this form is based on the hypothesis that the observation is not noisy or distorted (binary elements), as well as that a trade performed by a trader only reflects her opinion rather than a combination of opinions (right-invertibility). The complementary mapping for unobserved variables is given by a likewise defined matrix $F(t)$ such that $\sigma(t) = F(t) y(t)$.

Given the observations $s(t)$, our goal is to reconstruct the coupling matrix $J$, to infer the value of the parameters $b$ and $h$, as well as to estimate the missing values $\sigma(t)$. This can be done using a method relying on a Mean Field approximation and an Expectation-Maximization procedure, described in Campajola et al. \cite{campajola2019inference}. We give here a brief explanation of the method and refer to the original article for further details. The log-likelihood for the model is thus rewritten in terms of observed and unobserved variables
\begin{equation}
    \ell [\lbrace s \rbrace \vert J, h, b] = \log \mathrm{Tr}_{\sigma} \prod_t p \left[ \lbrace s(t+1), \sigma(t+1) \rbrace \vert \lbrace s(t), \sigma(t) \rbrace \right]
    \label{eq:loglik}
\end{equation}

The trace operator is computationally intractable for large systems with many missing values, so an approximation is needed in order to be able to compute the log-likelihood. The solution is given by the Martin-Siggia-Rose path-integral formalism \cite{msr1973} which, introducing a set of auxiliary variables, allows to shift the computational complexity from a cumbersome sum to a high-dimensional integral, for which approximate solution methods are known.

The result is an approximated version of the log-likelihood, function of the model parameters $\lbrace J,h,b \rbrace$ and of the posterior averages on unobserved opinions $\sigma_i(t)$, given as solutions of a set of \textit{self-consistency equations}
\begin{equation}
    m_a(t) = \mathbb{E}\left[\sigma_a(t)\right] = f_a[m(t)]
    \label{eq:selfcon}
\end{equation}
where the expectation is performed under the measure $p\left[ \lbrace \sigma \rbrace \vert \lbrace s, J, h, b \rbrace \right]$ and the $f_a[m(t)]$ are the non-linear self-consistency functions for the averages.

Calling
\begin{eqnarray*}
g_i(t) = \sum_j J_{ij}^{OO}(t) s_j(t) + \sum_b J_{ib}^{OH}(t) m_b(t) + h_i + b_i r(t) \\
g_a(t) = \sum_j J_{aj}^{HO}(t) s_j(t) + \sum_b J_{ab}^{HH}(t) m_b(t) + h_a + b_a r(t) \\
\end{eqnarray*}
where $J^{OO}(t) = G(t+1)JG(t)$, $J^{OH}(t) = G(t+1)JF(t)$ and so on, the approximated log-likelihood reads

\begin{align}
\mathcal{L} = \sum_t \Bigg[ \sum_i [s_i(t+1) g_i(t) - \log 2 \cosh(g_i(t))] &+ \nonumber \\
+ \sum_a [m_a(t+1) g_a(t) - \log 2 \cosh(g_a(t))] &+ \nonumber \\
 - \sum_a \left[\frac{1+m}{2} \log(\left(\frac{1 + m}{2}\right) + \frac{1 - m}{2} \log\left(\frac{1 - m}{2} \right)\right] &+ \nonumber \\
  - \frac{1}{2} \sum_i \left[ (1 - \tanh^2 (g_i(t)) \sum_b [J_{ib}^{OH}(t)]^2 (1 - m_b^2(t)) \right] &+ \nonumber \\
 - \frac{1}{2} \sum_a \left[(m_a^2(t+1) - \tanh^2(g_a(t)) \sum_b [J_{ab}^{HH}(t)]^2 (1 - m_b^2(t)) \right]& \Bigg]
\end{align}

while the self-consistency functions read

\begin{alignat}{2}
m_a(t) = \tanh \Bigg[& g_a(t-1) + m_a(t) \bigg[ &&\sum_{i} \left(1-\tanh^2 g_i(t+1) \right) [J^{OH}(t)]^2_{ia} + \nonumber \\ 
& &&+ \sum_{b} \left(m_b^2(t+1) - \tanh^2 g_b(t) \right) [J^{HH}(t)]^2_{ba} + \nonumber \\
& &&- \sum_{c}[J^{HH}(t-1)]^2_{ac} (1-m_c^2(t-1)) \bigg] + \nonumber \\
&+ \sum_{i} (s_i (t+1) -&& \tanh g_i(t) ) J^{OH}_{ia} + \nonumber \\
&+ \sum_{b} (m_b (t+1) -&& \tanh g_b(t) ) J^{HH}_{ba} + \nonumber \\
&+ \sum_i \frac{\tanh g_i(t)}{\cosh^2 g_i(t)} && \sum_{b} J^{OH}_{ib} (1 - m_b^2(t+1)) J^{OH}_{ia} + \nonumber \\
&+ \sum_c \frac{\tanh g_c(t)}{\cosh^2 g_c(t)} && \sum_{b} J^{HH}_{cb} (1-m_b^2(t+1) ) J^{HH}_{ca} \Bigg]  \label{selfcon::3}
\end{alignat}

Given the non-linearity of the $f_a$ functions the solution to these equations is found by iteration of the map until convergence.

Using the equations above, the model inference is performed by iterating a Gradient Ascent step \cite{Nesterov2008} in the $\lbrace J, h.b \rbrace$ space and the solution of the self-consistency equations, a process that eventually converges to a Maximum Likelihood Estimate of the parameters $\lbrace J, h, b \rbrace$ as well as an estimator for the hidden opinions in the form of $\hat{\sigma}_a(t) = \mathrm{sign}[m_a(t)]$. Once this is obtained, a parameter selection scheme called Decimation \cite{decelle2015inference} prunes all the irrelevant elements of $J$, resulting in a sparse directed and weighted network of interactions.

The Decimation procedure takes advantage of a transformation of the log-likelihood function in order to highlight when the removal of a parameter significantly impacts the quality of the whole model, by comparing the variation in log-likelihood and a linear interpolation between the complete model and a model with no parameters. Calling $\mathcal{L}_{max}$ the value of the log-likelihood coming from the optimization, and call $x$ the fraction of elements of $J$ that are pruned from the model (i.e. set to 0). Call $\mathcal{L}(x)$ the log-likelihood of the model with the fraction $x$ of pruned parameters and $\mathcal{L}_1$ the one with $x=1$, then one can define a Tilted log-likelihood as

\begin{equation*}
\mathcal{L}^{tilted}(x) = \mathcal{L}(x) - ((1 - x) \mathcal{L}_{max} + x \mathcal{L}_1)
\end{equation*}

Maximization of this transformed log-likelihood leads to the decimated model, which can then be interpreted as a sparse, directed and weighted network of lead-lag relationships.

We choose Decimation rather than the more popular LASSO parameter selection method because extensive numerical simulations show that the former performs better than the latter (see \cite{campajola2019inference} for more details).

Numerically the method requires $\mathcal{O}(N^2 \times T)$ operations at each iteration of the EM algorithm, meaning that its computational complexity grows quadratically with the number of variables one needs to model. This can become a problem when $N$ is in the order of thousands, but in our experience this approach can be adopted without resorting to high performance computing facilities if $N$ is in the range of several hundreds.

\section{Dataset}\label{sec:data}

Our dataset consists of all the trades performed in the period going from January 2012 to December 2013 on the eFX platform of a major dealer in the EUR/USD spot exchange rate market, including an anonymized identifier of the market agent requesting the trade, the volume and sign of the transaction, the time of request, and the price in EUR/USD quote offered by the dealer.

We select trades occurring on working days between 8AM and 4PM GMT and we split the dataset by month, resulting in 24 time series of trades with information about time, volume, sign, and identity of the counterpart. We then aggregate trades performed by the same agent $i$ within 5 minutes time windows and take the sign of the aggregate volume $V_i(t)$ of EUR acquired in exchange for USD as the information on whether the agent has sold ($V_i(t) < 0$), bought ($V_i(t) > 0$) or has stayed idle ($V_i(t) = 0$) at time $t$. Finally, we call $p_i$ the fraction of time intervals in which trader $i$ was active - that is, the fraction of non-missing data - and for each month we remove traders that were active in a fraction $p_i \leq 0.3$ of the total number of samples. 

The final dataset involves a total of 68 traders, with an average of 16 traders active each month, a minimum of 9 and a maximum of 29 and we report some statistics in Table \ref{tab:basestats}. To better understand the heterogeneity in the activity of market agents involved, we compute the Gini coefficient on the monthly $p_i$s and find the distribution of observations to be mostly homogeneous, typically having only one agent that is much more active than all the others.

% Thu Jan 24 12:58:13 2019
\begin{table}[ht]
\centering
\begin{tabular}{rrrrr}
  \hline
 & T & N & $\overline{s}$ & $\overline{p_i}$  \\ 
  \hline
Minimum & 679 & 9 & 0.02 & 0.45  \\ 
  Maximum & 2231 & 29 & 0.13 & 0.55 \\ 
  Mean & 2039.33 & 16.46 & 0.08 & 0.49 \\ 
  Stdev & 308.57 & 4.59 & 0.03 & 0.02 \\ 
   \hline
   \hline
 & Trader $p_i$ & $p_i$ Gini & Trader ACF1 & Flow ACF1 \\ 
  \hline
Minimum & 0.30 & 0.16 & -0.15 & 0.04 \\ 
  Maximum & 0.99 & 0.23 & 0.57 & 0.19 \\ 
  Mean & 0.49 & 0.19 & 0.07 & 0.12 \\ 
  Stdev & 0.18 & 0.02 & 0.07 & 0.05 \\ 
   \hline
\end{tabular}
\caption{Basic statistics of the dataset: (top) number of time steps $T$ and of agents $N$ of the monthly time series, monthly average sign of observed trades $\overline{s}$, monthly fraction of observations $\overline{p_i}$; (bottom) single trader monthly fraction of non-missing values $p_i$, monthly Gini coefficient of $p_i$, ACF at lag 1 of single traders and of the aggregate order flow.}\label{tab:basestats}
\end{table}

The sign of the aggregate trade volume $s_i(t) = \mathrm{sign}[V_i(t)]$ is intended as a proxy of the opinion the trader has at that time on whether the price should go up or down in the near future, while the zeros are intended as missing observations on their opinion. As shown in Table \ref{tab:basestats} the AutoCorrelation Function (ACF) at lag 1 on the aggregate order flow is typically higher than the average ACF of single traders, suggesting that traders act in coordination on a short lag, having their opinions diffuse gradually over a network of information spreading.

We infer the parameters of the Kinetic Ising Model of Eq. \ref{eq:KIM} on monthly subsets of data to account for non-stationarity and for traders that enter and exit the platform throughout the considered two year period. The outcome is a series of weighted and directed networks whose weighted adjacency matrix for month $k$ is $A(k) = J^T(k)$ (transposing conforms the matrix to the standard definition of adjacency matrix, which has non-zero element $a_{ij}$ if there is a link from node $i$ to node $j$), where the nodes are traders and the links represent an influence relation between the sign of the opinion of the origin node at time $t$ and the sign of the opinion of the end node at time $t+1$. The links can have either positive or negative weight: when it is positive it means that the follower tends to agree with the leader opinion, while when it is negative they tend to disagree.

Since we hypothesize that returns $r(t)$ can affect the trading behaviour of traders, we introduce the 5-minutes log-returns as a control variable in the model, with a trader-specific parameter $b_i$ capturing their reaction to a price change in the previous time window. We use the mid-price in the order book of the EBS electronic inter-dealer exchange to which the dealer has access as a market member: although traders do not specifically trade at that price, it is the only price indicator that we can reliably use while not introducing trade-specific effects.

\section{Results}\label{sec:results}

\begin{figure}[t]
    \centering
    \includegraphics[width=.7\textwidth]{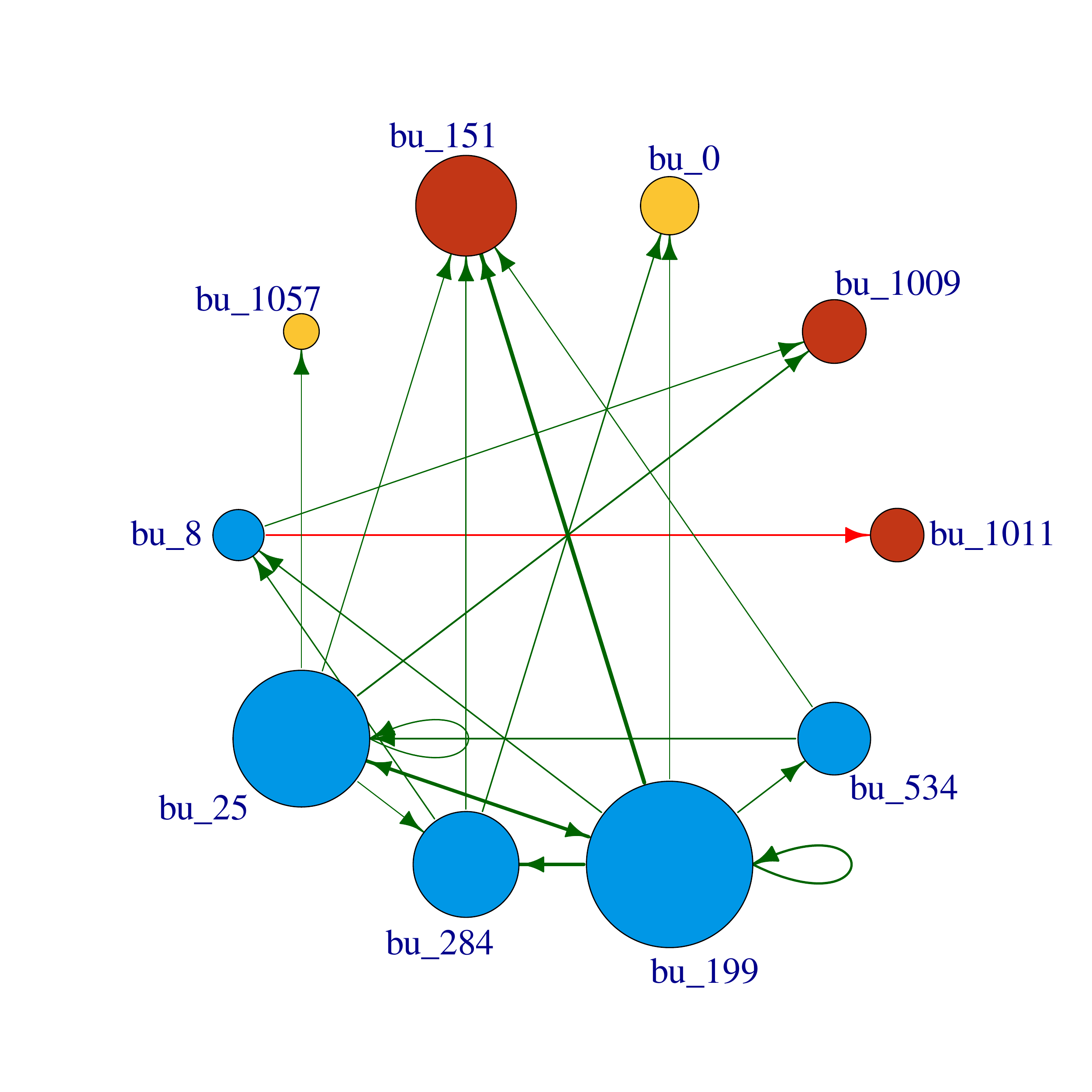}
    \caption{The inferred lead-lag network at month 13. Node coloring follows the PageRank influence categorization described in section \ref{sec:influencers}, the size of both nodes and links is proportional to their strength and weight, respectively. The link color indicates whether it is positively (green) or negatively (red) weighted.}
    \label{fig:sample_net}
\end{figure}

The resulting networks (as for example the one shown in Figure \ref{fig:sample_net}) are then analysed to find out whether there are traders that are more influential than others, how the network changes over time, and how accurate is the prediction of trade signs. In Section \ref{sec:influencers} we define a characterization of the nodes as influencers and followers based on an adapted weighted version of the PageRank \cite{brin1998anatomy} measure as proposed by Kiss and Bichler \cite{kiss2008identification}. Then in Section \ref{sec:neighb} we compute the persistence of the neighbourhood of nodes as described by Nicosia et al. \cite{nicosia2013graph} to quantify the local stability of the networks in time and try to disentangle degree-related effects from preferential attachment by comparing the results with the ones obtained by randomly rewiring the networks and reshuffling the time series.
In Section \ref{sec:OOS} we compute the out-of-sample accuracy of prediction of trade signs to evaluate model performance compared to a Logistic AutoRegressive (LAR) model of order 1, taking as input the previous trade sign of trader $i$ (where available) and the last log-return. We also evaluate the forecasting and nowcasting performance of the model, utilizing parameters fitted on one month to predict trade signs in the next one, always comparing with the LAR benchmark.
Finally in Section \ref{sec:GC} we show a further interesting feature of our approach which allows us to define a micro-level herding measure. We take this measure and run a Granger Causality analysis between it and a set of liquidity imbalance measures computed on the order book of the EBS inter-dealer exchange to highlight the functioning of the multi-dealer market and emphasize the role of the dealer as a liquidity provider.

\subsection{Influence network: key players and properties}\label{sec:influencers}

\begin{figure}[ht!]
    \centering
    \includegraphics[width=1\textwidth]{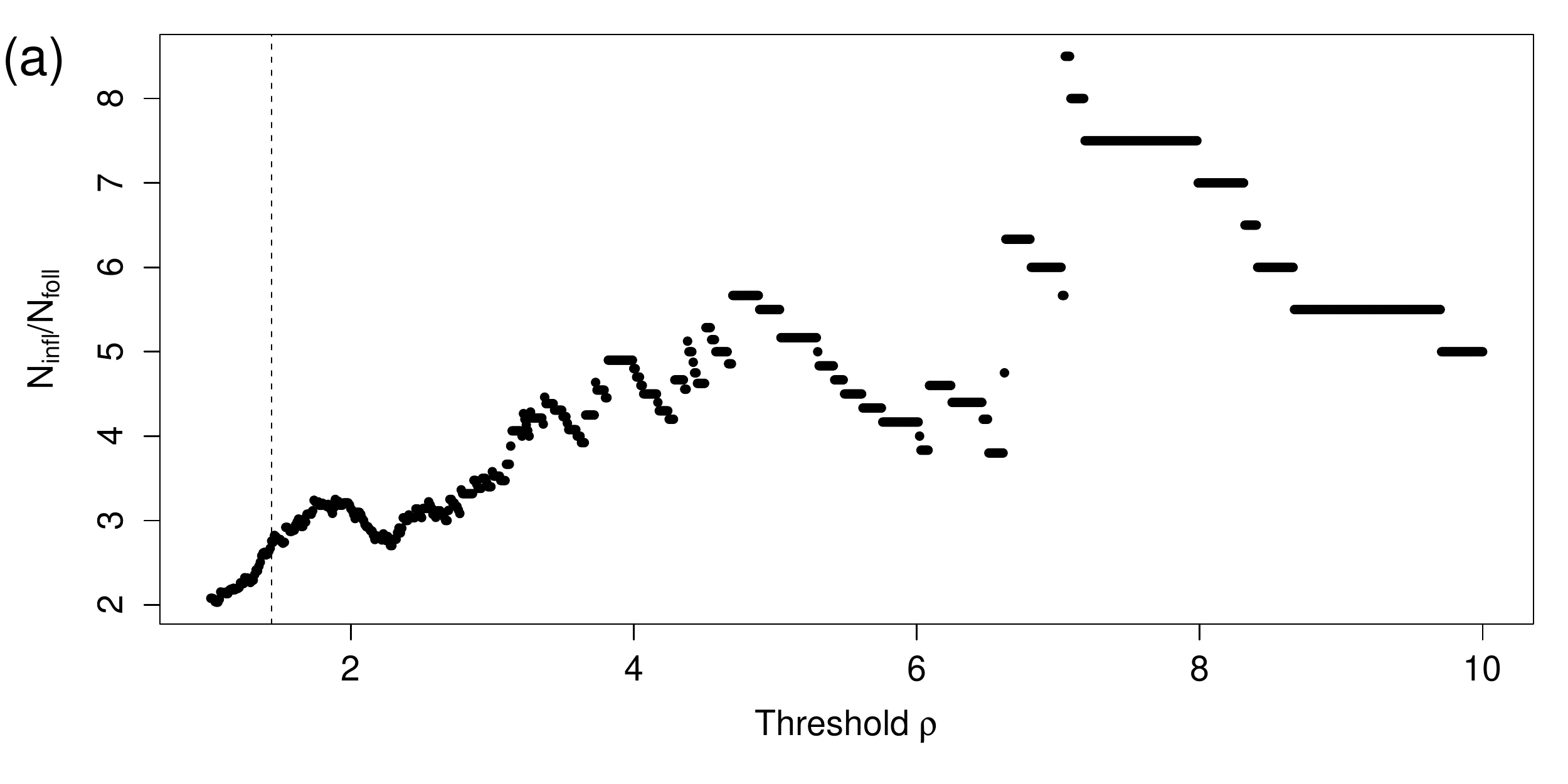}
    \includegraphics[width=1\textwidth]{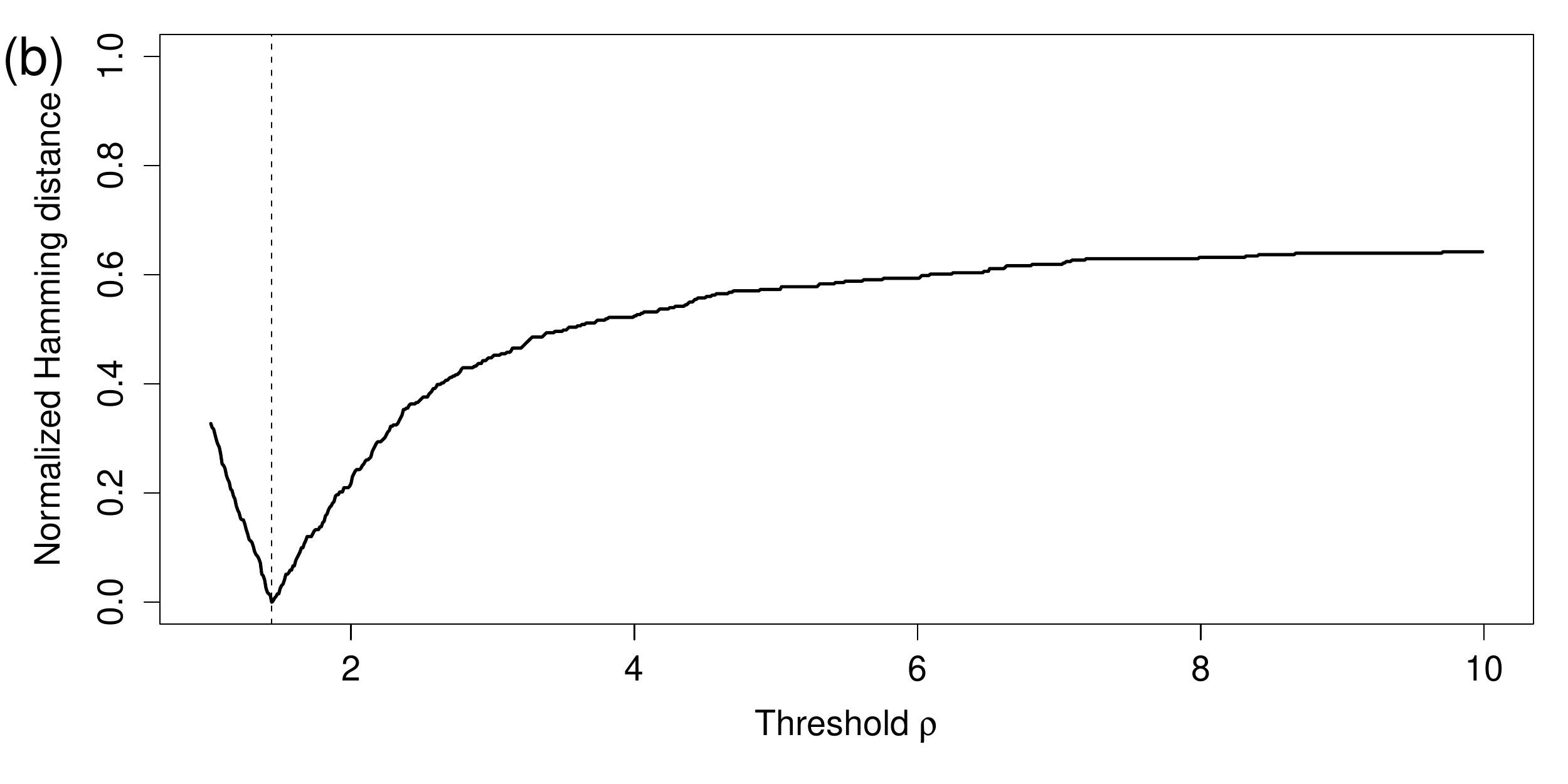}
    \caption{Stability of the PageRank ratio categorization. (a) Ratio between the number of identified influencers and the number of identified followers varying the threshold parameter $\rho$; (b) Normalized Hamming distance between the categorization chosen optimizing Eq. \ref{eq:optimrho} and other categorizations varying $\rho$. Vertical dashed lines mark the value of $\rho^*$.}
    \label{fig:PRstab}
\end{figure}

It is our interest to identify key actors in the market that carry information about behavioral trends and that might ``lead the pack", forerunning the order flow. To categorize traders in our network we adopt the measure developed by Kiss and Bichler \cite{kiss2008identification} as a modification of the PageRank algorithm by Brin and Page \cite{brin1998anatomy}. The PageRank measure identifies important nodes based on how likely it is that a so-called \textit{random surfer}, that is a random walker with some probability of restarting from a random node, ends up on some specific node of the network. In particular we want to label our nodes in 3 categories: influencers, followers, and neutrals. Kiss and Bichler define the Weighted PageRank (WPR) and the Weighted SenderRank (WSR) measures, where a node has higher WPR (WSR) the larger the relative strength\footnote{The strenght of a node is the sum of the weights of all links pointing at (in-strength) or departing from (out-strength) that node.} of incoming (outgoing) links it has from (towards) highly ranked nodes. Both these measures have a minimum value related to a parameter $f$ called \textit{damping factor}, representing the probability that the random surfer keeps walking instead of jumping to a random node, which we choose to be the literature standard $0.85$ for both. The resulting measure for node $i$ is then defined as

\begin{eqnarray*}
\mathrm{WSR}_i = (1 - f) + f \sum_{j \in L_i} \frac{w_{ij}}{S_i} \mathrm{WSR}_j \\
\mathrm{WPR}_i = (1 - f) + f \sum_{j \, s.t. \, i \in L_j} \frac{w_{ji}}{S_j} \mathrm{WPR}_j
\end{eqnarray*}

where $L_i$ is the set of nodes that have an incoming link from node $i$, $w_{ij}$ is the weight of the link between $i$ and $j$ and $S_i = \sum_{L_i} w_{ij}$ is the out-strength of node $i$.

Notice that since the links can have negative weights we take the absolute value of the weight to account for negative influence as well. We then define the category $C_i^t$ of trader $i$ in month $t$ based on the ratio between their WSR and WPR:
\begin{equation}
C_i^t = 
\begin{cases}
\mathrm{Influencer}, \, \mathcal{I} & \text{if}\ \mathrm{WSR}_i^t/\mathrm{WPR}_i^t > \rho \\%\mathrm{WSR}_i > 2m \\
\mathrm{Follower}, \, \mathcal{F} & \text{if}\ \mathrm{WSR}_i^t/\mathrm{WPR}_i^t < 1/\rho \\%\mathrm{WPR}_i > 2m \\
\mathrm{Neutral}, \, \mathcal{N} & \text{otherwise}
\end{cases}\label{eq:PRcatrule}
\end{equation}
where $\rho$ is a threshold ratio that can be arbitrarily decided. To make this decision less arbitrary, we try to find an optimal value of the ratio in order to maximize categorization diversity cross-sectionally while keeping it consistent through time. The idea is thus to minimize a measure of diversity for the single agent across months, while maximizing the same measure between different agents in the same month, and taking $\rho$ as the optimal in terms of Euclidean distance from the ideal case of perfect trader consistency in time and perfect uniformity of categorization cross-sectionally.

Call $p_i^{\rho}(C) = 1/T \sum_t \delta (C_i^t, C)$ the empirical probability at which trader $i$ is assigned to category $C$ using threshold $\rho$: the measure of diversity we choose is the normalized \textit{Total Variation Distance} $d(p_i^\rho)$ from the uniform distribution, namely
\begin{equation*}
    d(p_i^{\rho}) = \frac{3}{2} \sup_{C \in \lbrace \mathcal{I}, \mathcal{F}, \mathcal{N} \rbrace} \left| p_i^{\rho}(C) - \frac{1}{3} \right|
\end{equation*}
where $p_i^{\rho}$ is compared to the uniform distribution which takes value $1/3$ for all categories, and the factor $3/2$ is making sure that $d(p_i^\rho)$ is normalized to $1$ in the case of maximum homogeneity, while it is $0$ at maximum diversity. Call $\check{C}_i^{\rho} = \arg \max p_i^{\rho}(C)$ the most frequent categorization of trader $i$ at threshold $\rho$, and define the frequency of category $\check{C}$ among the $\check{C}_i$s as $f^\rho(\check{C}) = 1/N \sum_i \delta(\check{C_i}, \check{C})$. Finally, call $\zeta^\rho = d(f^{\rho})$ the cross-sectional diversity between the most frequent categories the traders are assigned to. Then we optimize $\rho$ as
\begin{equation}
    \rho^* = \arg \min_{\rho} \left[ (\zeta^\rho)^2 + (\mathbb{E}_i[d(p_i^\rho)] - 1)^2 \right]
    \label{eq:optimrho}
\end{equation}
that is we minimize the Euclidean distance from the ideal case of having each trader in the same category every month ($\mathbb{E}_i[d(p_i^\rho)] = 1$) and evenly spread categories across agents ($\zeta^\rho = 0$), obtaining a threshold value of $1.44$.

\begin{figure}[ht!]
\includegraphics[width=1\linewidth]{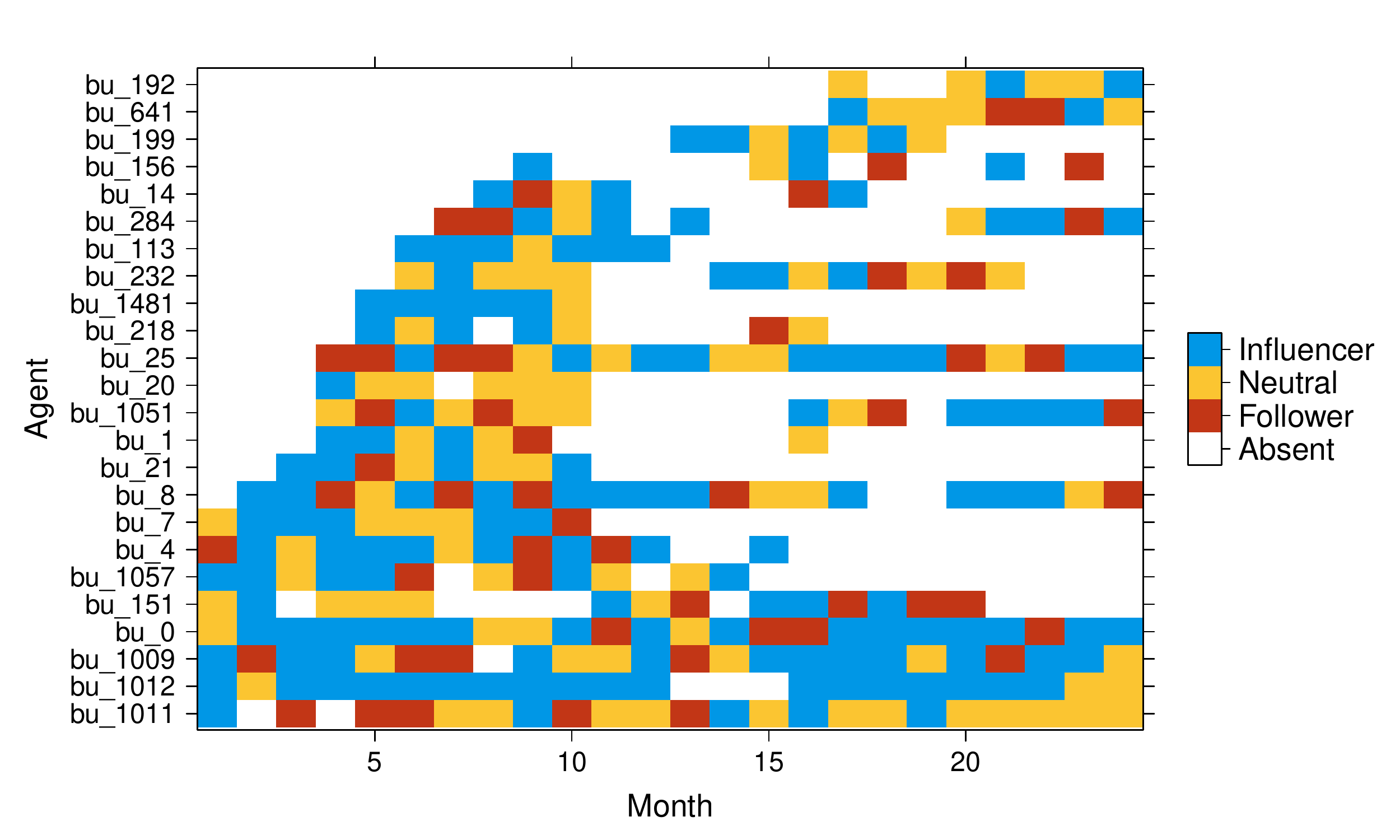}
\caption{PageRank categorization of agents across months.}\label{fig:pagerank}
\end{figure}

We show how the selection of the threshold affects the categorization in Fig. \ref{fig:PRstab}, plotting the influencers/followers ratio and the Hamming distance between the chosen and all other categorizations. In the region surrounding the chosen threshold the ratio of influencers to followers is rather stable at around $3$, and the normalized Hamming distance (that is the fraction of categories changing between two choices of $\rho$) between the chosen category and its neighbourhood is rather low and smoothly varying when moving away from the chosen threshold, a sign that the categorization is stable enough to justify using this selection method.
  
The resulting categories for traders that exist in the data for more than 5 months are shown in Fig. \ref{fig:pagerank}. It is rather interesting to see how some traders show a consistent behaviour across the whole dataset being identified mostly as influencers (see for example trader \#1012, \#113 and \#1481), while others have a more swinging nature.

\subsection{Network persistence}\label{sec:neighb}

\begin{figure}[t]
\includegraphics[width=1\linewidth]{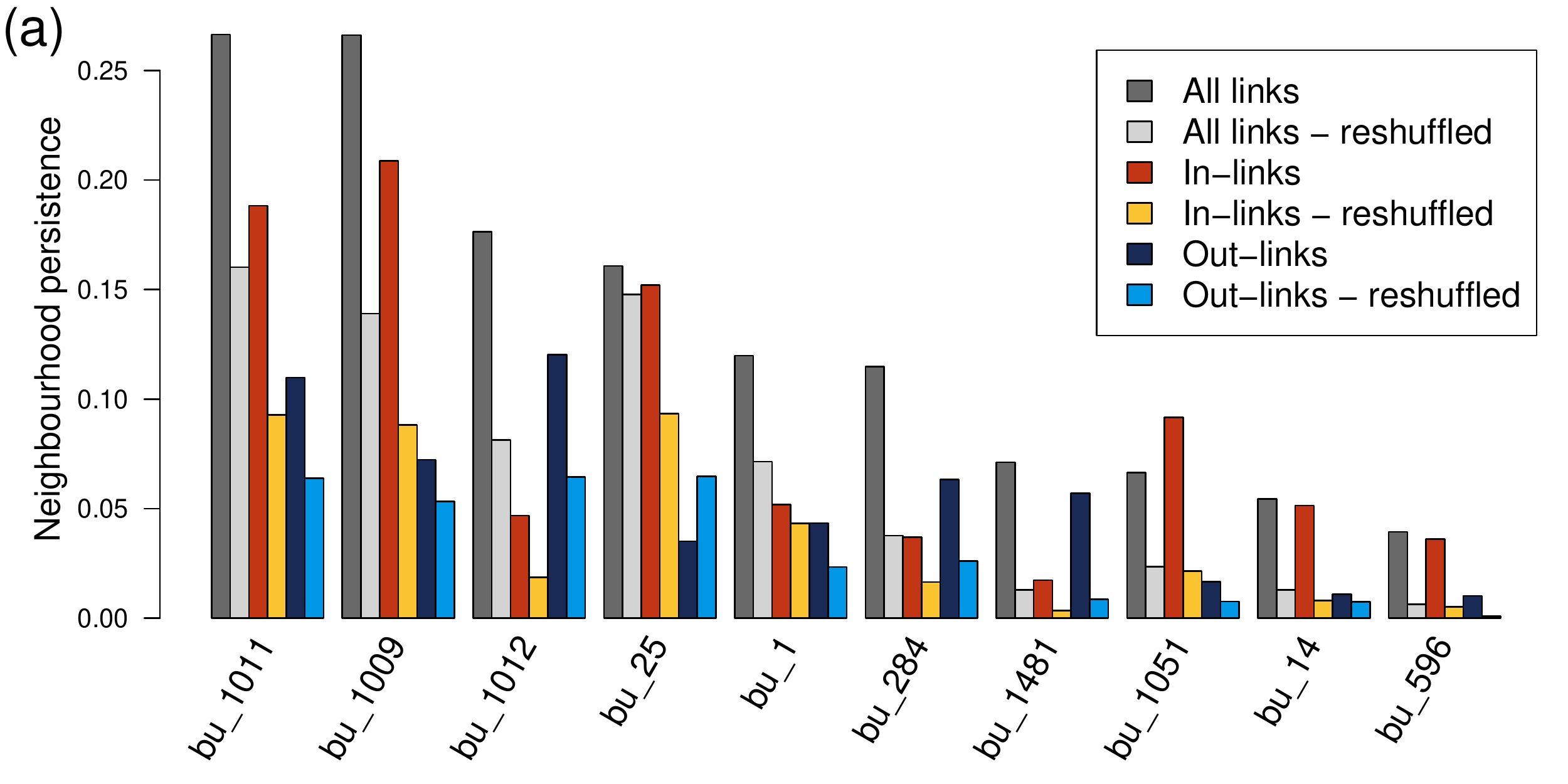}
\includegraphics[width=1\linewidth]{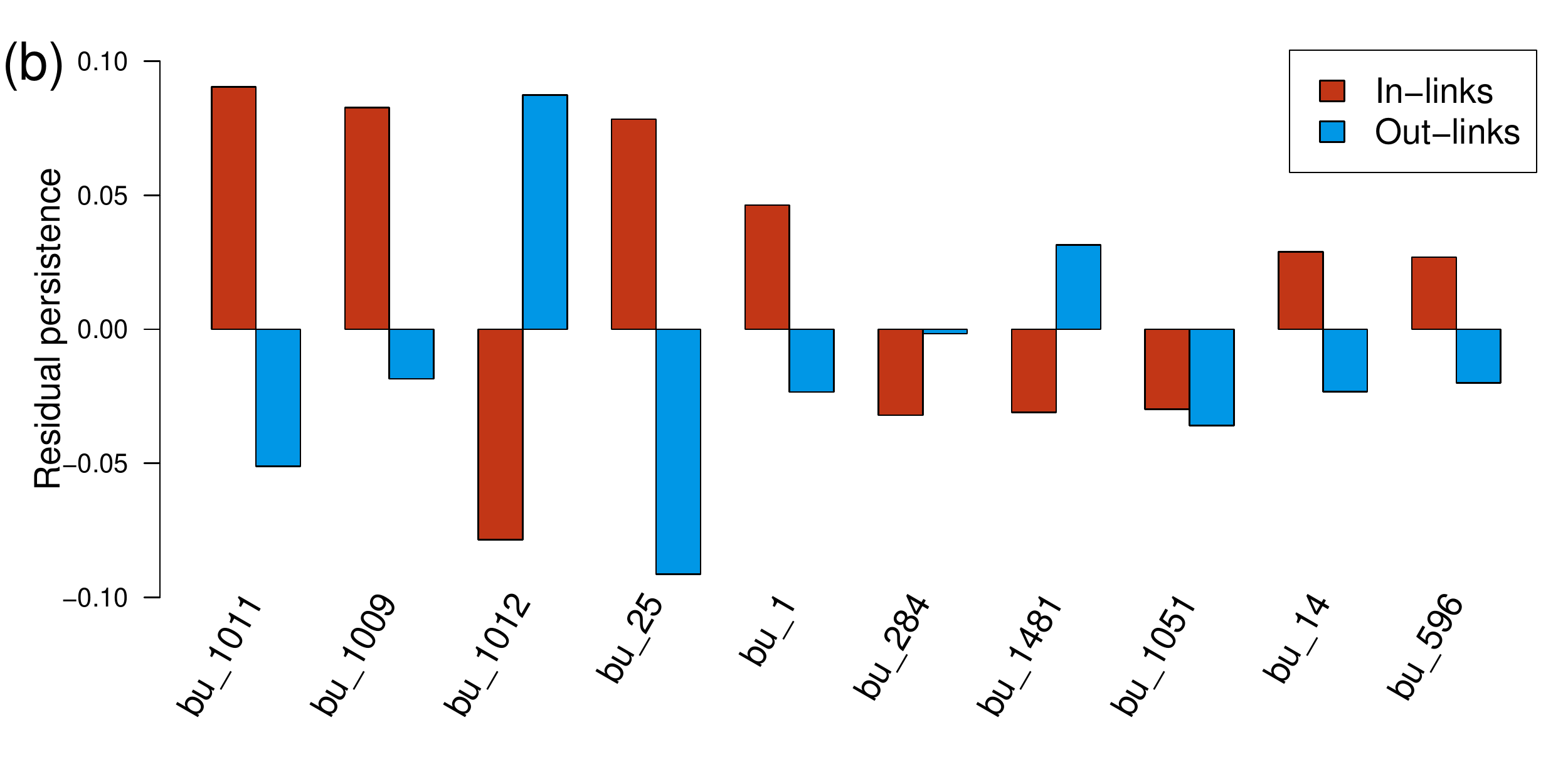}
\caption{(a) Neighbourhood persistence measured before and after a random reshuffle of the network time series for a subset of nodes; (b) Residual in- and out-neighbourhood persistence after a degree-preserving rewiring of the network for the same subset of nodes.}\label{fig:persist}
\end{figure}

To understand how variable the network is from month to month we compute the neighbourhood persistence measure proposed by Nicosia et al. \cite{nicosia2013graph}, defined as

\begin{equation}
D_i (t, t+1) = \frac{\sum_j a_{ij}(t)a_{ij}(t+1)}{\sqrt{\sum_j a_{ij}(t) \sum_k a_{ik} (t+1)}}
\label{eq:persistence}
\end{equation}
where $a_{ij}$ are the elements of the network adjacency matrix. Since the network is directed we compute the measure on the three possible neighbourhoods - the in, out, and total neighbourhood - changing the summation indices appropriately: in particular, Eq. \ref{eq:persistence} refers to the out-neighborhood, while summing over rows instead of columns produces the measure for the in-neighborhood and the total is obtained by using the symmetrized adjacency matrix $A^T + A$. We compare it to the same measure averaged over $10,000$ order randomizations of the network time series to isolate the actual persistence in time from the average connectivity the trader has. In Figure \ref{fig:persist}a we plot the two quantities for the $10$ nodes in the network that show the largest persistence and for all neighbourhood types. We see that these nodes tend to have abnormally persistent neighbourhoods, sometimes more in the in-neighbourhood and sometimes in the out-neighbourhood, a sign that some preferential relationships exist and replicate themselves in time. We also test the persistence by doing a random degree-preserving rewiring of the networks (while keeping the temporal structure) in order to remove the effect of the node degree, as a more connected node is more likely to have a more persistent neighbourhood than a less connected one. Figure \ref{fig:persist}b shows the difference in the directed neighbourhood persistence between the original networks and the rewired ones. When this residual persistence is positive it means that the node has a persistence higher than in the null configuration model and viceversa.

The results show that there are indeed nodes that show a higher (or lower) persistence in their neighbourhoods even when ruling out the effect of the in- and out-degree, while this is typically not true for the undirected version (which roughly corresponds to the sum of the two). A node with a higher persistence of the out-neighbourhood is a node that attaches preferentially to some other nodes, meaning, in our convention, that it has influence over a persistent set of nodes, while the opposite is true for a node with higher persistence of the in-neighbourhood. For example, node \#1011 has overly persistent incoming links and non-persistent outgoing links, meaning it is typically influenced by the same set of nodes, while its influenced neighbours are more randomly selected. The opposite happens for node \#1012, which is indeed consistently recognized as an influencer by Weighted PageRank. Overall this analysis shows that, even if the network density is rather high and it is difficult to extract significant community structures, there is evidence of some preferential attachment mechanism at work in the directed network.

\subsection{Out of sample validation and forecasting}\label{sec:OOS}

\begin{figure}[t]
\includegraphics[width=1\linewidth]{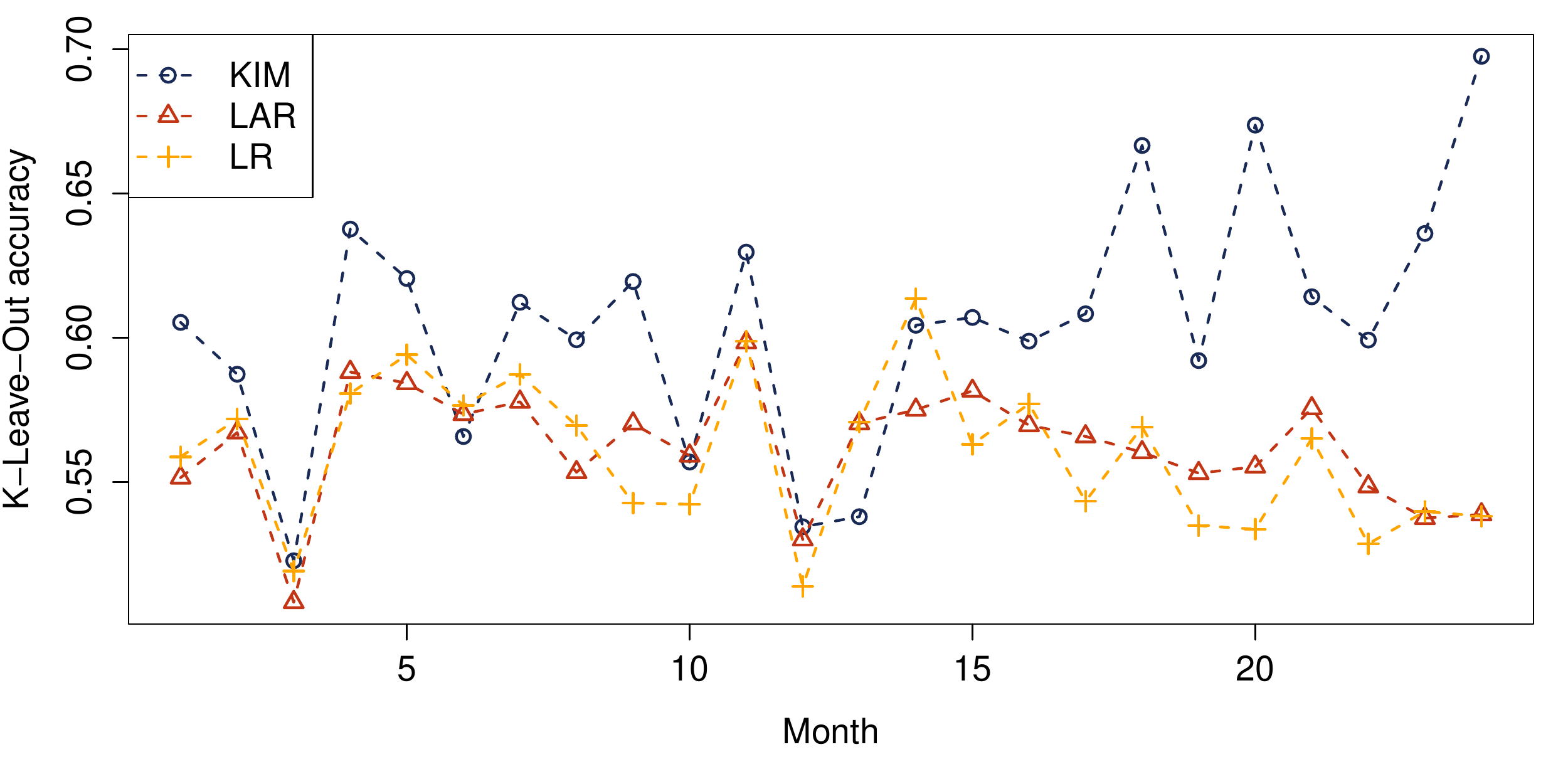}
\caption{Out-of-sample K-leaveout accuracy of the KIM model compared to a Logistic AutoRegressive (LAR) model and to a Logistic Regression on log-returns (LR) for every month in the dataset.} \label{fig:perf}
\end{figure}

In this subsection we perform out of sample validation and forecasting of the presented model. Specifically, we neglect some observed trades and we test whether our model is able to correctly guess them. In the forecasting exercise we instead train the model in a subperiod and test whether we are able to predict the trading activity in the following subperiod. In all cases we consider models with and without 5-minutes log-returns as a control variable.

%We compare models obtained with and without accounting for the exchange rate changes, since we hypothesize that returns can affect the trading behaviour of traders. We thus introduce the 5-minutes log-returns as a control variable in the model, with a trader-specific parameter capturing their reaction to a price change in the previous time window. We use the mid-price in the order book of the EBS electronic inter-dealer exchange: although traders don't have direct access to that market, it is the only price indicator that we can reliably use at this frequency not introducing trade-specific effects on the price.

In Figure \ref{fig:perf} we plot the performance one has predicting out-of-sample trade signs using the Kinetic Ising Model compared to the average of $N$ logistic univariate logistic regressions, with the log-returns as an independent variable and, in the AutoRegressive version (LAR), the trade sign of the trader at the previous time interval (if available). The performance is measured by K-leaveout cross-validation, consisting of hiding $5\%$ observations from the sample and then comparing the predicted trade sign with the actual one. The measure is then the fraction of correctly identified trade signs.% While in some cases there is a performance gain taking the returns into account, it appears that in most of the months it is not particularly relevant or even detrimental for out of sample validation, albeit by a few percent. This happens despite the cross-correlation function between the 5-minutes returns and the average trade sign is significantly different from $0$ (in particular it is negative).

Overall the performance of our model is better than the benchmarks in a range from $5\%$ to over $10\%$ (excluding a couple of months where it does slightly worse), and in the best case the model predicts trade signs with $70\%$ accuracy, while on average it scores around $60\%$. While being nothing too extraordinary, this result shows that the model can provide a valid platform for descriptive and forecasting purposes.

The difference in performance between the KIM and the univariate logistic regression models is larger when the cross-correlation at lag 1 between the order flow and the log-returns is non-significant (not shown here). This tells us that, while the simpler models capture what is probably the most important interaction observed in the market (the reaction of traders to price changes), when this interaction is weaker they fail to capture any significant effect. However a significant amount of coordination persists regardless of whether it is caused by price movements or by other mechanisms, and it can be explained by our modelling approach.

%\begin{figure}[t]
%	\centering
%	\includegraphics[width=1\linewidth]{fig7a.pdf}
%	\includegraphics[width=1\linewidth]{fig7b.pdf}
%	\caption{One-step-ahead forecasting performance, using previous month parameters. (a) Forecasting accuracy averaged over the first week of the following month, using all traders, only the influencer subset, or a LAR model. (b) K-Leave-Out performance within the same month for reference, as well as the three days, one week and one month ahead average performance.}
%	\label{fig:forecast1}
%\end{figure}

We thus try to use the Kinetic Ising Model to forecast order signs: as a proof of concept, we take the result from one month and use the inferred parameters to produce the one-step-ahead forecasts in the next month. 

Calling $\lbrace J^{M}, h^M, b^M \rbrace$ the set of inferred model parameters at month $M$ and $s^M_i(t)$ the observed order sign of trader $i$ at time $t$ in month $M$, we forecast one step ahead using $\hat{s}_i^M(t+1) = \mathrm{sign}(\check{m}_i(t+1))$, where

\begin{equation*}
	\check{m}^M_i(t+1) = \tanh \left[ h^{M-1}_i + b^{M-1}_i r_{t} + \sum_{j \in \mathrm{obs}(t)} J^{M-1}_{ij} s^M_j(t) + \sum_{b \notin \mathrm{obs}(t)} J^{M-1}_{ib} m^M_b(t) \right]
\end{equation*}
and $\mathrm{obs}(t)$ is the set of observed indices at time $t$. This quantity is then compared to the time $t+1$ observations and the average number of correct guesses is reported as the forecasting performance. Notice that every time an observation is added the $\check{m}^M(t)$ vector is updated through Eq. \ref{eq:selfcon} to include the new information and keep the forecasting just to one step ahead of the observations.

We analyze the performance of the KIM when using all traders or only the influencers subset to predict future order signs compared to the same task performed with a LAR model. The results (not reported for the sake of space) show that there is no significant increase in performance by introducing the multivariate modeling, and restricting the prediction to using only traders that were identified as influencers in the previous month doesn't seem to change radically the forecasting accuracy. 
%In Figure \ref{fig:forecast1}b we compare performances measured within the same month by the K-Leave-Out cross-validation accuracy with the one-step-ahead forecast accuracy averaged over the next three days, week and month.
We observe that the performance is marginally better ($\sim 55 \%$) than a random guess and that is rather stable across time horizons (we also tested the one-step ahead forecasts using models several months after their inference without noticing significant changes). Our hypothesis is that both the LAR and the KIM methods, when used for forecasting, mostly rely on the log-returns to guess the next trade, which we believe is the reason why the accuracy of predictions is just a few percents higher than a coin flip and it does not vanish at longer time horizons. While this may seem at odds with the results shown in the previous sections, it has to be pointed out that the main objective of our model is to infer the state of investors when they do not trade, not forecasting, and that to do so we take advantage of future information in Eq. \ref{eq:selfcon}, something that is clearly not possible for one step ahead forecasts.

\subsection{Predicting liquidity from inferred opinions}\label{sec:GC}

\begin{figure}[t]
    \includegraphics[width=1\linewidth]{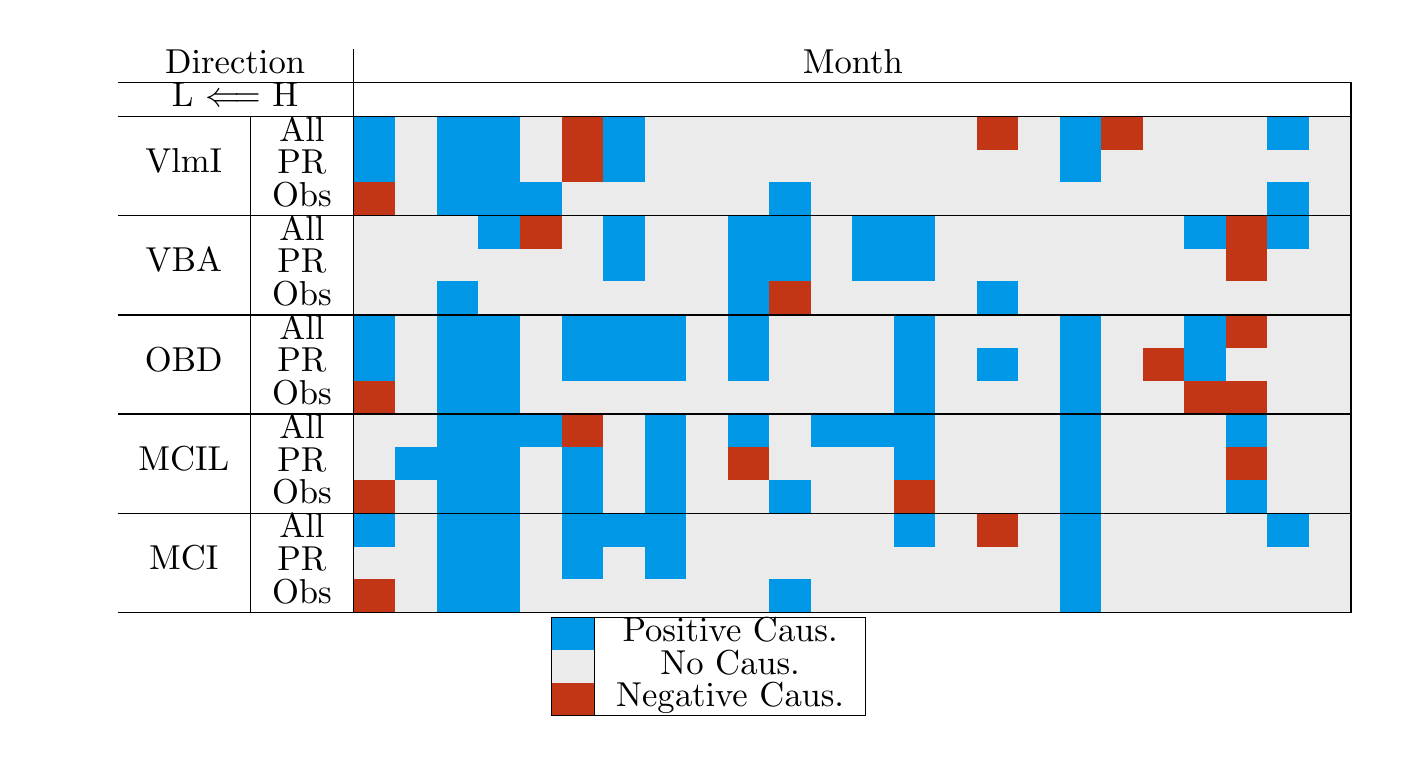}
    \caption{Causality relations from herding to liquidity. The model is based on 5 minutes lags with order up to 12, that is one hour, and the reported sign is the one of the coefficient is the one for the minimum order showing Granger Causality effects. The herding measure either accounts for all traders (``All"), only the ones belonging to the influencers group under the PR influence measure (``PR") or only the observed ones (``Obs").}
    \label{fig:GC_HerdLiq}
\end{figure}

One possible use for our modelling approach is to produce a ``herding" measure, given by the average opinion of traders at any point in time. Indeed a by-product of the model estimation is a maximum likelihood estimate of the unobserved opinions in the market, which we can use to generalize the buy-sell imbalance that trade signs show to an implied opinion imbalance. Typically herding is defined as an irrational behaviour that crowds show where a large fraction of agents co-ordinate based on social interaction rather than as a reaction to information, often resulting in unjustified macroscopic phenomena as, in the case of financial markets, price volatility jumps and dramatic liquidity imbalances. Herding has been documented in fund industry \cite{grinblatt0} as well as in institutional and individual investors \cite{nofsinger,grinblatt1}, and in market members \cite{vaglica}.

The herding measure we define, as a simplification of the one already present in \cite{lakonishok}, is
\begin{equation*}
    H(t) = \frac{1}{N}\sum_{i=1}^N\hat{y}_i(t)
\end{equation*}
where $\hat{y}_i(t)$ is either the observed sign of the transaction $s_i(t)$ executed by trader $i$ at time $t$ or the one inferred as $\mathrm{sign}[m_i(t)]$. The main difference with the existing definitions is of course that we include also inferred states.

\begin{figure}[t]
    \centering
    \includegraphics[width=1\linewidth]{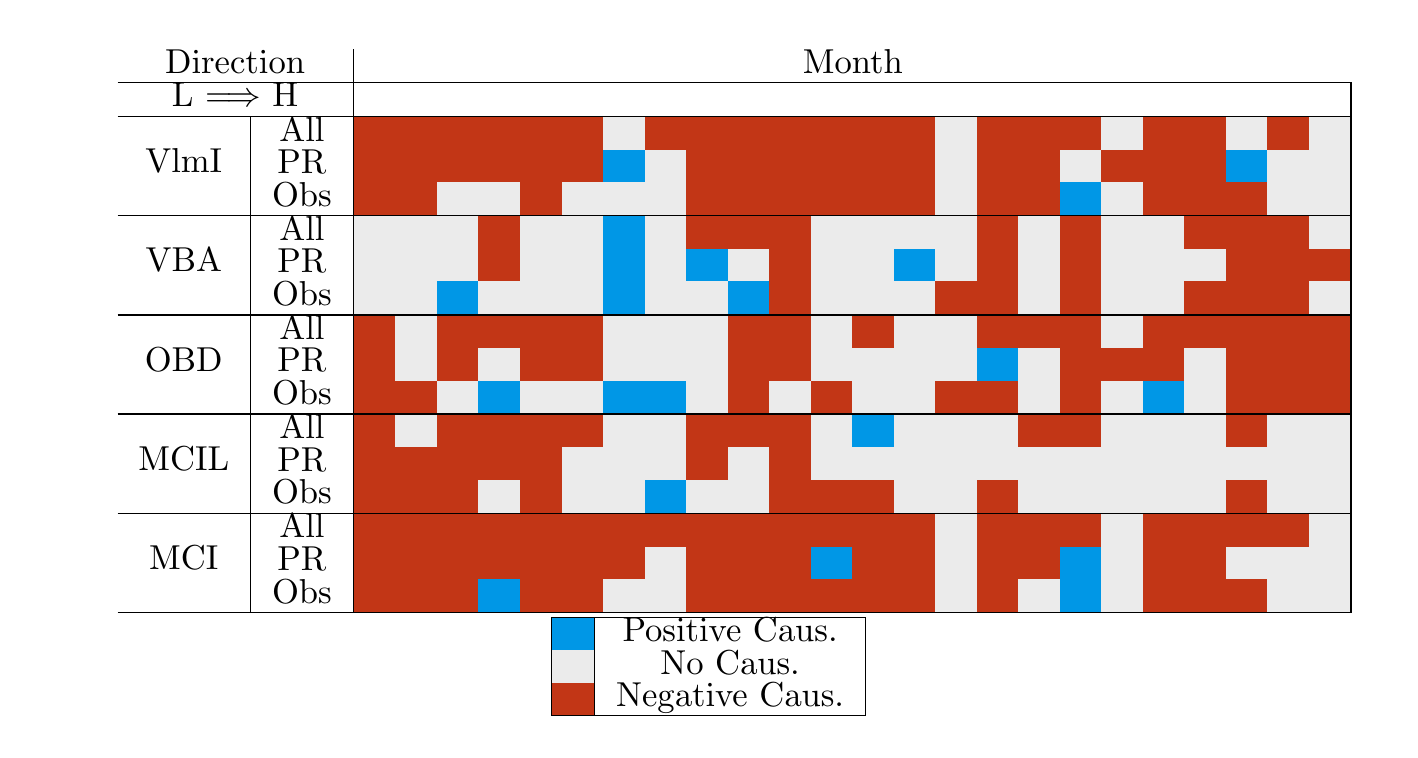}
    \caption{Causality relations from liquidity to herding. The model is based on 5 minutes lags with order up to 12, that is one hour, and the reported sign is the one of the coefficient is the one for the minimum order showing Granger Causality effects. The herding measure either accounts for all traders (``All"), only the ones belonging to the influencers group under the PR influence measure (and ``PR") or only the observed ones (``Obs").}
    \label{fig:GC_LiqHerd}
\end{figure}

We want to show how this measure can be used to study a typical problem dealers are confronted with, that is facing poor liquidity conditions in the inter-dealer market when their inventory becomes unbalanced due to unexpected trading pressure from clients. To this end, we take into account a set of liquidity imbalance measures in the interdealer market:

\begin{itemize}
\item \textbf{VBA}: Dollar Volume at best Bid-Ask. It is the difference between the volume of limit orders at the best bid level and the volume of limit orders at the best ask, normalized by the total volume at those levels;
\item \textbf{OBD}: Order Book Depth. It is the difference between the number of levels that have to be explored to execute a buy market order of $10^7$ units of currency (which is the typical imbalance that the dealer accumulates in a 5-minutes time window) and an equal sell market order size;
\item \textbf{MCI imbalance}: It is the imbalance between the Marginal Cost of Immediacy between the ask and bid side. MCI, introduced by Cenesizoglu et al. \cite{cenesizoglu2018bid}, is defined as
\begin{align*}
 \mathrm{MCI}_A &= \frac{\mathrm{VWAPM}_A}{\mathrm{Vlm}_A}\\
 \mathrm{VWAPM}_A &= \log \frac{\frac{\mathrm{Vlm}_A}{\sum_{l=1}^{L} Q_{A,l}}}{0.5 (P_{A,1}+P_{B,1})} \\
 \mathrm{Vlm}_A &= \sum_{l=1}^{L} P_{A,l}Q_{A,l}
\end{align*}
where $P_{A,l}$ is the price at level $l$ on the Ask side and $Q_{A,l}$ is the quantity available at level $l$ on the Ask side. The same can be defined for the Bid side and the measure we use is $MCI_A - MCI_B$.
The quantity is computed for $L=10$ (MCI) and for $L=OBD$ (MCIL), in order to capture book-wide imbalances as opposed to typical transaction size imbalances.
\item \textbf{VlmI}: Dollar Volume Imbalance. It is the normalized amount of dollars in orders on the bid side of the book minus the same quantity on the ask side;
\end{itemize}

All these measures are defined such that a positive imbalance means that liquidity is higher for the bid side of the order book, that is it is easier for a market participant to execute a sell market order (the asset is always considered to be EUR and the quotes are given in USD, as in the dealer platform data).

To explore the relationship between these measures and $H(t)$ we run Granger Causality tests on pairwise Vector AutoRegressive (VAR) models for which we report results in Figures \ref{fig:GC_HerdLiq} and \ref{fig:GC_LiqHerd}. We choose this method over alternatives such as Transfer Entropy \cite{bossomaier2016introduction, novelli2019large} for its simplicity and ease of interpretation, but it is possible that comparing with more elaborate techniques can produce more interesting results. The figures show the causality relations we find in both directions and specify whether the coefficient of the VAR model for which the causality is found is positive or negative. If a positive (negative) causality is found, it means that an increase in the first variable is causing an increase (decrease) in the other. To reduce the number of false positives we implement the false discovery rate (FDR) method \cite{benjamini1995controlling} for multiple hypothesis testing, setting the significance threshold at 0.05.

\begin{figure}[t]
    \centering
    \includegraphics[width=1\linewidth]{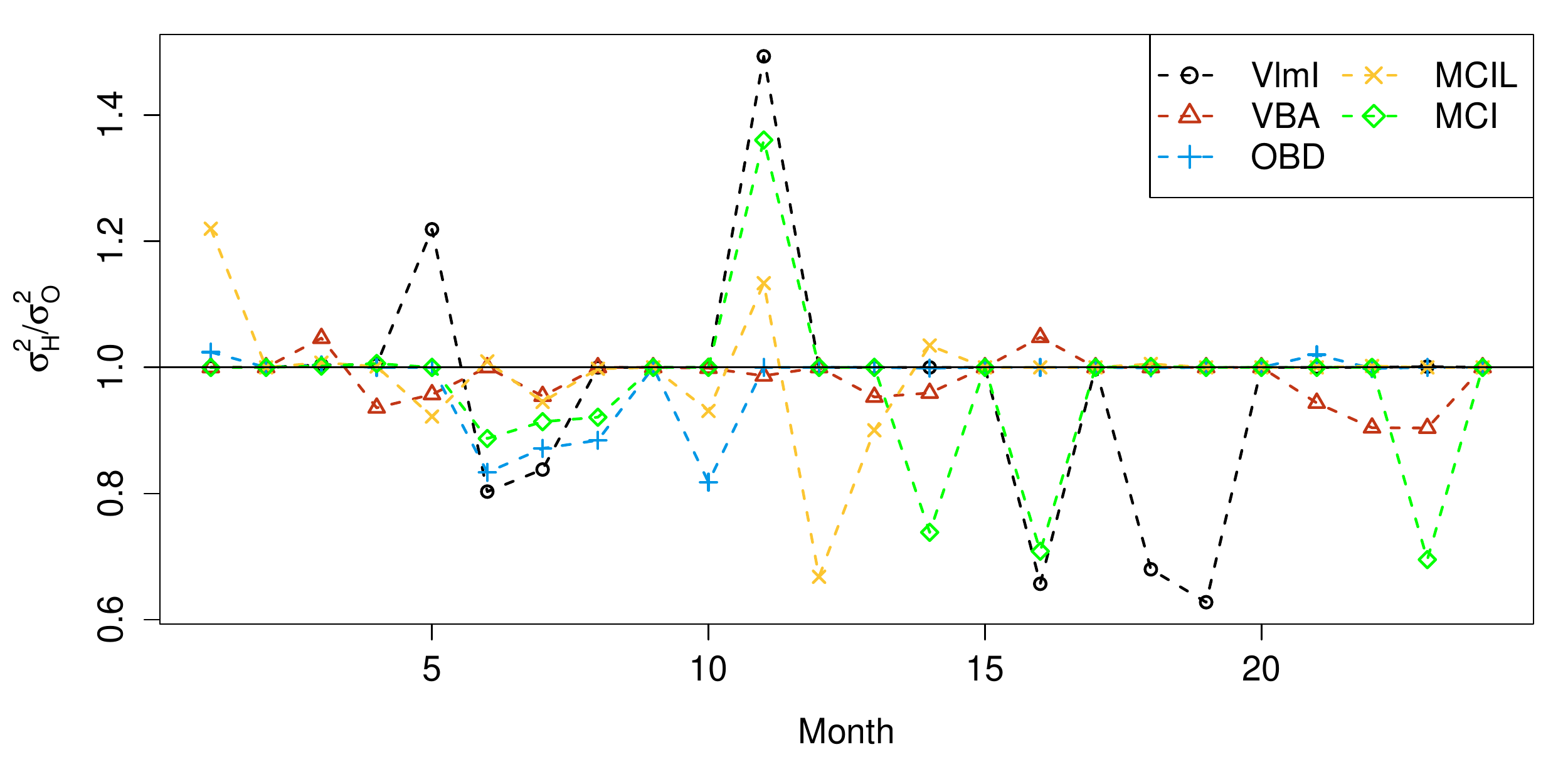}
    \caption{Residuals variance ratio between models using our herding measure $H_t$ or the observed trade sign imbalance. The ratio is mostly less than 1, showing models have a better goodness of fit when using $H_t$.}
    \label{fig:GC_varrat}
\end{figure}

The results highlight the importance of the dealer in distributing liquidity and absorbing temporary imbalance in the supply and demand: indeed most of the relations running in the direction $H \rightarrow L$, that is Herding to Liquidity, are positively signed, while the opposite is true when looking at the $L \rightarrow H$ direction. This means that when the herding measure is positive, and so the majority of traders on the eFX platform is buying EUR, the liquidity on the EBS market will make it harder for the dealer to quickly rebalance her inventory as the imbalances are typically positive, meaning it is easier to sell than to buy EUR. On the other hand, when the EBS market conditions are favorable for the dealer to sell (positive L), this is typically followed by a majority of traders selling EUR to the dealer (negative H), as it is likely that the dealer is offering better quotes given the ease she has in unloading excess inventory.

We also show how the herding relation to liquidity is typically unchanged whether one includes in its computation all traders or only the subset of influencers as identified by the Weighted PageRank measure, meaning that they are indeed among the most informative traders in this sense, while only using the observed trades and ignoring the opinions reconstructed through the Kinetic Ising Model one finds less and more incoherent causality relations. As a further argument in support, the quality of the Vector AutoRegressive model fit is generally better when considering our measure over the observed trades imbalance, as shown by Figure $\ref{fig:GC_varrat}$. There we compare the variance of the residuals on the liquidity side of the VAR model when using our herding measure $H(t)$ or the observed trades as the other model variable. We see that the ratio is typically less than 1, meaning the variance is smaller (and thus the fit better) with our measure.

\begin{figure}[t]
    \centering
    \includegraphics[width=1\linewidth]{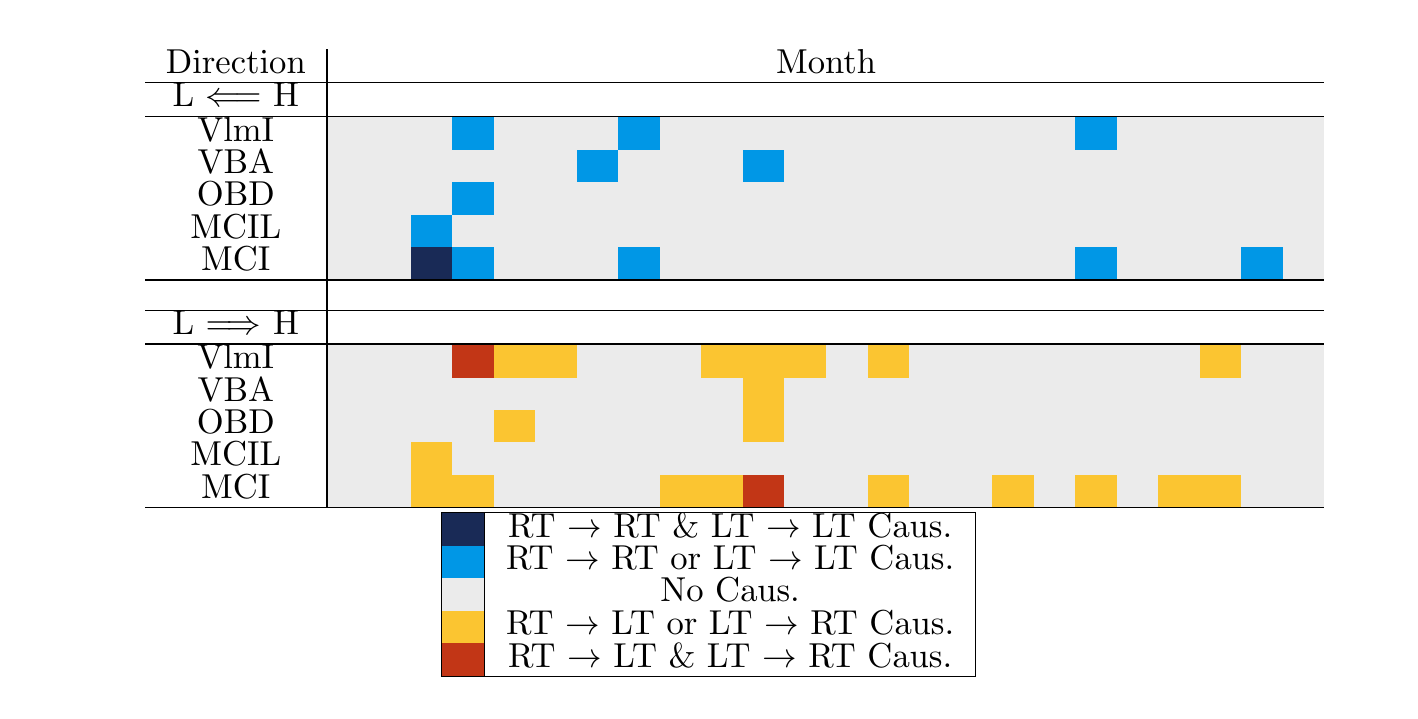}
    \caption{Tail Granger Causality relations identified by the test of Hong et al. \cite{hong2009granger}. We see how the results agree with the Granger Causality in mean, showing the strong connection between the markets also on extreme events.}
    \label{fig:tailgc}
\end{figure}

To further investigate this relation, we apply the test of Granger Causality in tail originally proposed by Hong et al. \cite{hong2009granger} between $H(t)$ and the liquidity measures. It is indeed interesting to see whether the Granger Causality only appears on average or it shows also between extreme events. The test is built to identify causality relations between binary time series, representing occurrences of extreme (tail) events with respect to recent history. Such events are identified as values of the liquidity or herding measure that exceed the 90\% empirical conditional quantile (or are below the 10\% quantile), measured as proposed by Davis \cite{davis2016verification} on the past 2 hours of data at all points. The measurements above the 90\% threshold are denominated as ``right tail" (RT) events, while the events below the 10\% one are ``left tail" (LT) events. We show the results of the analysis in Figure \ref{fig:tailgc}, where we see that the picture given by the Granger Causality in mean is confirmed and the effects are particularly recurrent for the total volume imbalance (VlmI) and the MCI measure.

This last result highlights why it could be important to estimate the unseen opinions of traders: given the multi-dealer structure of the spot rate FX market, a dealer only has a partial picture of what the supply and demand for the asset looks like at any given time, offered by the trades she sees from her clients. However these clients might have access to other dealer platforms and trade at their convenience with one or the other, thus hiding their opinion to the single dealer while still using market liquidity. This is then reflected in the order book of the inter-dealer market, where liquidity deteriorates whenever a shift in supply and demand occurs and makes it costly for the dealers to efficiently rebalance their inventories.

\section{Conclusions}\label{sec:concl}

In this article we have introduced the Kinetic Ising Model as a method to infer causal relationships between trader activities in a financial market at high frequency and to achieve a better estimate of the aggregate supply and demand at any point in time. We apply the model to a proprietary dataset offered to us by a major dealer, selecting the most active traders on their electronic foreign exchange trading platform to study the lead-lag relationships that occur among them and how their behaviour affects the state of liquidity on another market, the EBS inter-dealer electronic exchange. We show that several market players can be identified as influencers, that is they are typically leading the order flow on the 5 minutes time-scale, and that their trading activity and opinion explains liquidity imbalances on the EBS market. Studying the persistence in time of the network structure on the local scale, we notice that some nodes have directed neighbourhoods that replicate through months, an effect that further validates the inferred lead-lag relations and that matches quite well with the results from the influence analysis. We also test the forecasting performance that can be achieved with this model, finding that both the model and the LAR benchmark are not particularly well-suited for the purpose and the inclusion of the lead-lag relationships does not change the forecast significantly. We do not investigate the nature of these lead-lag relationships, but we propose they should be interpreted as the effect of different traders following similar strategies with different reaction times, leading to one or more traders consistently forerunning the others, rather than a more ``direct" type of influence caused by actual social interactions. Finally we defined a herding measure based on the inferred opinions, which we show has much stronger Granger Causality relations with the state of liquidity on the inter-dealer market than just observed trade signs, exposing a mechanism that highlights the role of the dealer in providing immediacy to her clients and absorbing the cost of liquidity.

\section*{Acknowledgements}
We thank Adam Majewski for his help in the initial steps of this project. CC acknowledges SNS for financial support of the project SNS18\_A\_TANTARI. DT acknowledges GNFM-INDAM for financial support.

\bibliographystyle{unsrt}%
\bibliography{biblio.bib}%

\begin{appendices}
\renewcommand\thefigure{\thesection.\arabic{figure}}
\renewcommand\theequation{\thesection.\arabic{equation}}
\setcounter{figure}{0}
\setcounter{equation}{0}
\section{Simulation study}

In this appendix we present a brief simulation study aiming to provide further insight to the reader regarding the results that should be expected from this approach. We produce a synthetic dataset of opinions based on the basic statistics of our data, summarized in Table \ref{tab:basestats}, by simulating the Kinetic Ising Model fixing $N$, $T$ and the distribution of traders probability of observation $p_i$ to closely resemble the ones that we observe in the data. We thus choose $N = 20$, $T = 2000$ and the distribution of $p_i$ is assumed to be a Beta distribution, $p_i \sim B(\alpha, \beta)$, with parameters $\alpha = \beta \approx 4.01$. The value of the parameters is obtained following Pham and Turkkan \cite{pham1992determination} in order to be consistent with the observed average Gini coefficient of $p_i$ and the mean cross-sectional $p_i$ of $0.5$. The remaining free parameters are the ones directly related to what we aim to infer, that is the structure of the interaction matrix $J$ and the magnitude of its elements.

We thus explore several degrees of sparsity of the underlying $J$ matrix by sampling it as an Erd\H{o}s-R\'{e}nyi random graph with parameter $d_J \in [0,1]$ describing the probability of a link, i.e. the density of the graph, and vary the parameter $J_1$ which regulates the magnitude of the coupling coefficients assuming that for the existing links $J_{ij} \sim \mathcal{N}(0, J_1 / \sqrt{N})$. The scaling with $\sqrt{N}$ is necessary to be able to compare parameters coming from models with different $N$, as it correctly normalizes the sum in Eq. \ref{eq:KIM}.

\begin{figure}
    \centering
    \includegraphics[width=1\linewidth]{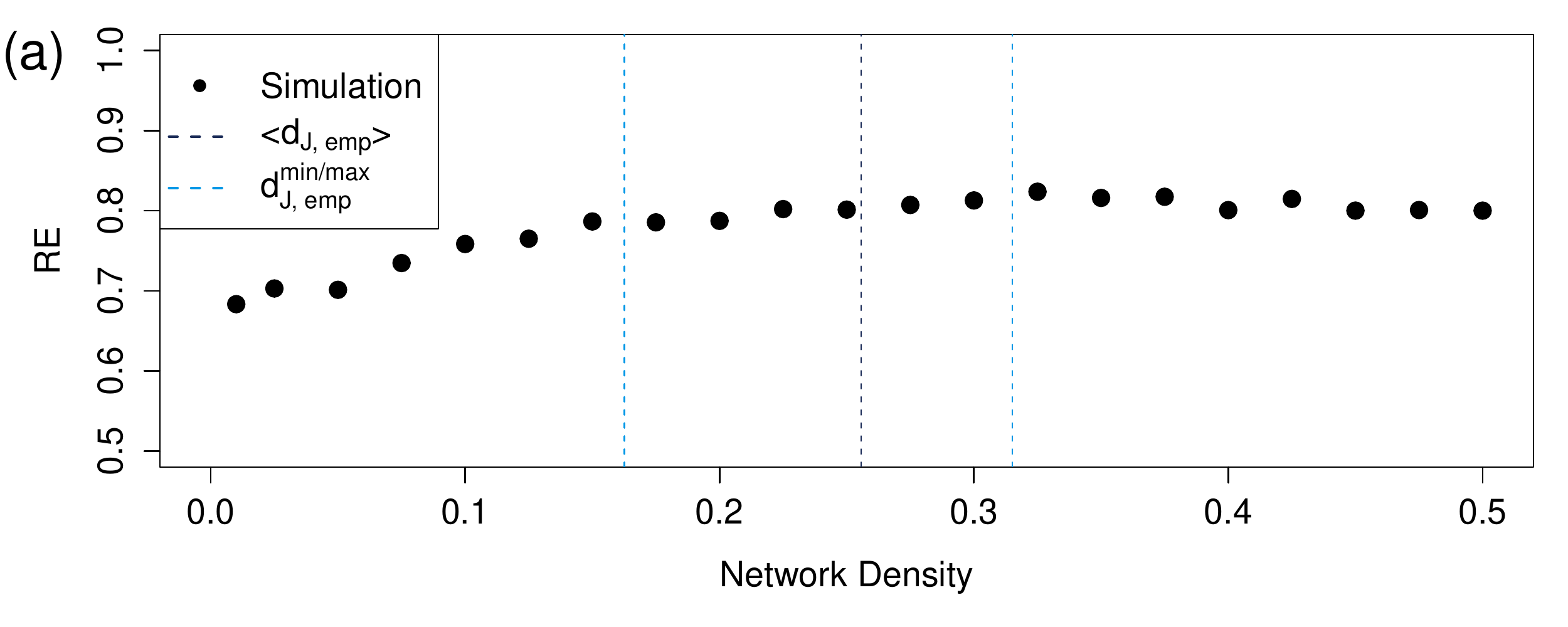}
    \includegraphics[width=1\linewidth]{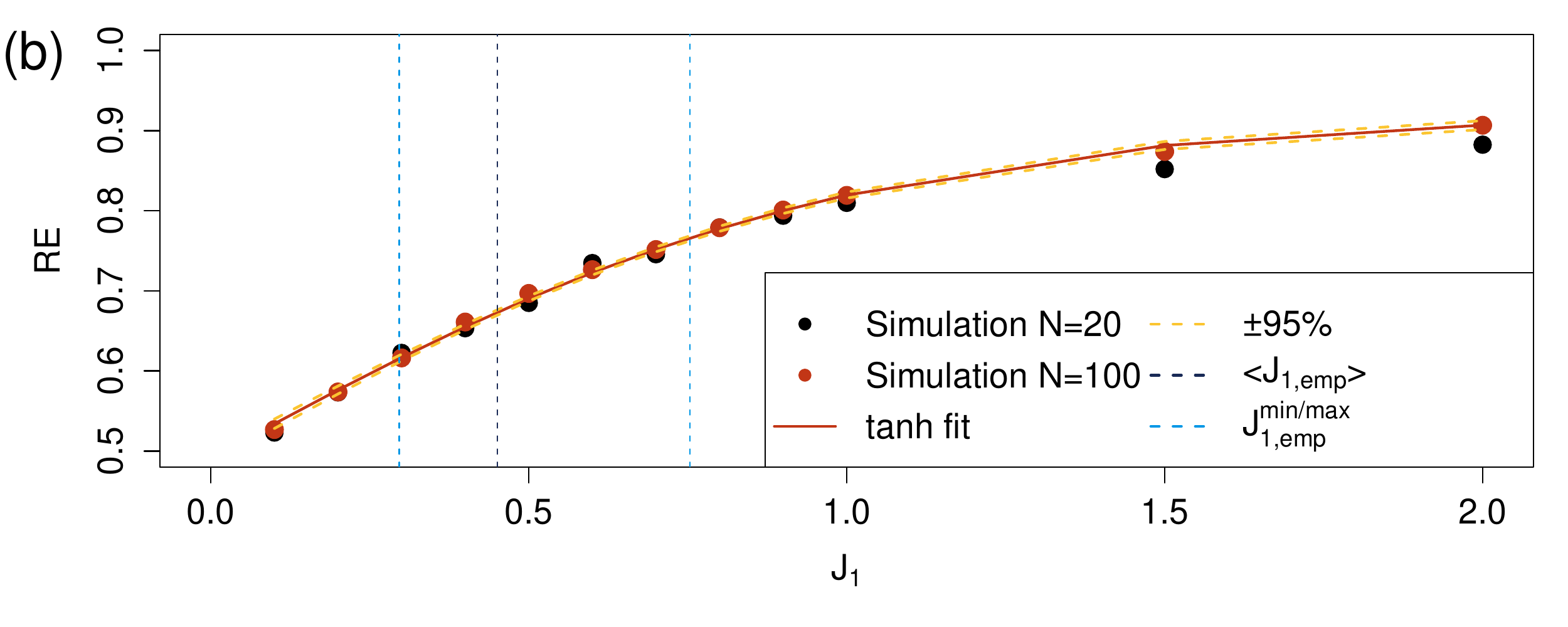}
    \includegraphics[width=1\linewidth]{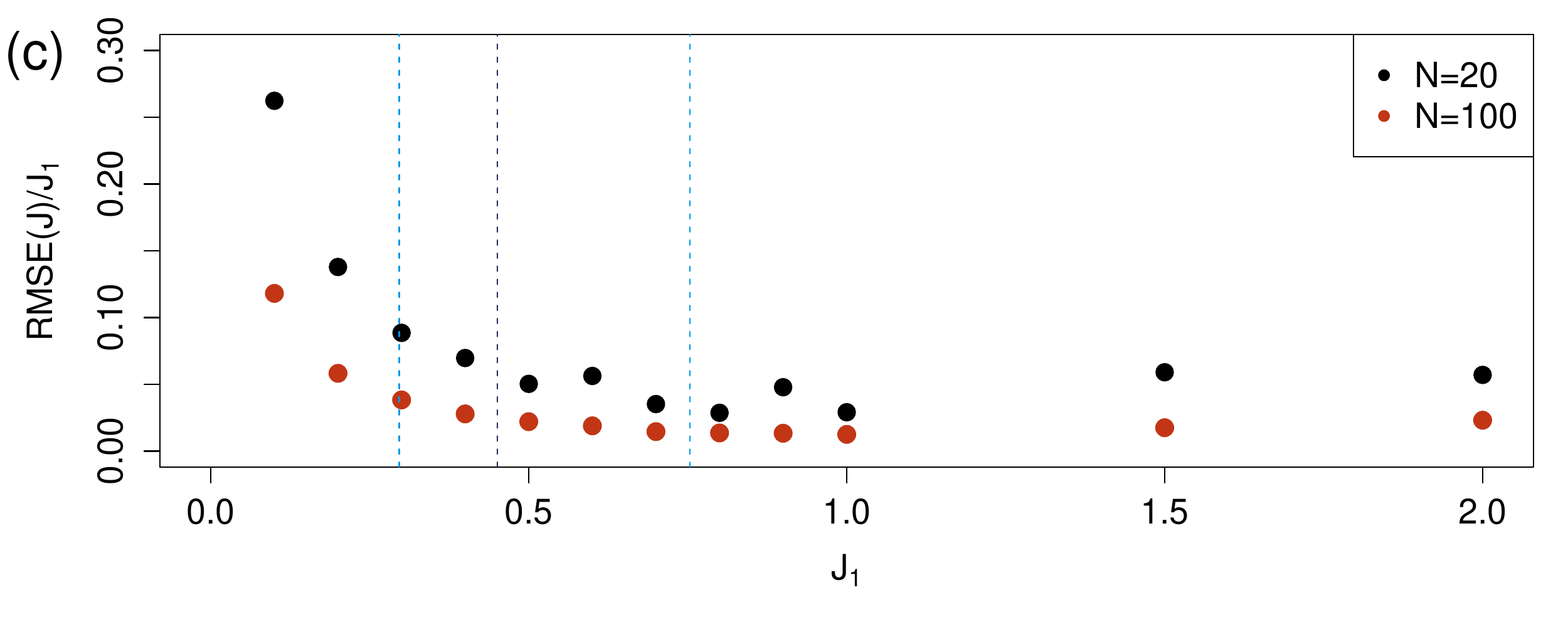}
    \caption{Results from the simulation study. (a) Reconstruction Efficiency (RE) varying the network density $d_J$. We see that, besides a slightly decreasing efficiency at very low densities, the expected performance is more or less constant. (b) RE varying the magnitude of couplings $J_1$. Here we see a clear relation between the two, highlighted by the hyperbolic tangent fit; (c) Rescaled Root Mean Squared Error (RMSE) on J elements. In all panels the blue lines show the region where the models inferred from trading data are situated.}
    \label{fig:simstudy}
\end{figure}

We show these results in Figure \ref{fig:simstudy}, by plotting the Reconstruction Efficiency (RE) of hidden opinions, that is the fraction of hidden opinions that is correctly guessed, varying $d_J$ and $J_1$ and showing the region we find empirically from our trading dataset. While no particular dependence of the RE is to be expected from the network density, as shown in Figure \ref{fig:simstudy}a, in Figure \ref{fig:simstudy}b we also see how it is instead strongly dependent on the magnitude of the couplings. This is also predictable from the theory, as we show with an hyperbolic tangent fit. Indeed the probability distribution of a hidden value $\sigma_i$ given $m_i$ is 
\begin{equation}% \tag{A.1}
    p(\sigma_i(t) = \pm 1 \vert m_i) = \frac{1 \pm m_i}{2}
\end{equation}

Here $m_i$ depends from $J_1$ through Eq. \ref{selfcon::3} where $J_1$ is, given its definition and the Central Limit Theorem, the typical size of any sum of the kind $\sum_j J_{ij} s_j$ or $\sum_b J_{ib} m_b$, assuming all $m_b$s are estimated with no error and $N \rightarrow \infty$. Indeed the coefficients of the fit $\mathrm{RE} = a + b\tanh(J_1)$ are found to be $a = 0.49 \pm 0.03$ and $b = 0.43 \pm 0.03$ to $95 \%$ confidence for $N=20$, $T=2000$ and similar results are obtained for a larger system with $N=100$, $T=10000$. The small discrepancy between the theoretical value of $b=0.5$ and the one we measure in simulations is most likely due to the presence of more than one hidden value, introducing uncertainty in the estimation of $m$ itself.

We also plot the Root Mean Squared Error on $J$ elements in relative terms to the magnitude of the parameters $J_1$, showing that in the region in which we find our inferred parameters there is a RMSE of roughly $5\%$ in simulations, giving an idea of the error one could expect on the estimates.

Of course this is an ideal case, where the data generating process and the model coincide, meaning that these results have to be interpreted as upper bounds in performance. We indeed see that our out-of-sample performance results of Figure \ref{fig:perf} are below the RE we get from simulations, but we argue that they are not that far from those given the size of the inferred parameters, meaning that even if the model is more than likely misspecified and an oversimplification of reality it still captures significant features from the data.

\end{appendices}

\end{document}